\documentclass[12pt]{article}
\usepackage{amssymb,amsmath}
\usepackage[affil-it]{authblk}
\usepackage[top=0.75in, bottom=0.75in, left=0.75in, right=0.75in, dvips]{geometry}
\usepackage{caption}
\begin{document}
\textwidth 10.0in
\textheight 9.0in
\topmargin -0.60in
\title{Renormalization Scheme Dependence  \\
 with Renormalization Group Summation}

\author{D.G.C. McKeon \thanks{dgmckeo2@uwo.ca}}
\affil {Department of Applied Mathematics, University of Western Ontario, London, ON N6A 5B7, Canada\\
and \\ 
Department of Mathematics and
Computer Science, Algoma University, Sault St.Marie, ON P6A
2G4, Canada}

\maketitle

\maketitle

\begin{abstract}
We consider all perturbative radiative corrections to the total $e{^+}e{^-}$ annihilation cross section $R_{e{^+}e{^-}}$ showing how the renormalization group (RG) equation associated with the radiatively induced mass scale $\mu$ can be used to sum the logarithmic contributions in two ways.  First of all, one can sum leading-log (LL), next-to-leading-log (NLL) etc. contributions to $R_{e{^+}e{^-}}$ using in turn the one-loop, two-loop, etc. contributions to the RG function $\beta$. A second summation shows how all logarithmic corrections to $R_{e{^+}e{^-}}$ can be expressed entirely in terms of the log-independent contributions when one employs the full $\beta$-function.  Next, using Stevenson's characterization of any choice of renormalization scheme by use of the contributions to the $\beta$-function arising beyond two-loop order, we examine the RG scheme dependence in $R_{e{^+}e{^-}}$ when using the second way of summing logarithms.  The renormalization scheme invariants that arise are then related to the renormalization scheme invariants found by Stevenson.  We next consider two choices of renormalization scheme, one which can be used to express $R_{e{^+}e{^-}}$ solely in terms of two powers of a running coupling, the second which can be used to express $R_{e{^+}e{^-}}$ as an infinite series in the two-loop running coupling (i.e., a Lambert $W$-function).  In both cases,  $R_{e{^+}e{^-}}$ is expressed solely in terms of renormalization scheme invariant parameters that are to be computed by a perturbative evaluation of $R_{e{^+}e{^-}}$.  We then establish how in general the coupling constant arising in one renormalization scheme can be expressed as a power series of the coupling arising in any other scheme.  We then establish how by using a different renormalization mass scale at each order of perturbation theory, all renormalization scheme dependence can be absorbed into these mass scales when one uses the second way of summing logarithmic corrections to $R_{e{^+}e{^-}}$.  We then employ the approach to renormalization scheme dependency that we have applied to $R_{e{^+}e{^-}}$ to a RG summed expression for the Coleman-Weinberg effective potential $V$ in a massless scalar model with quartic self coupling, showing  that the previously derived result that $V$ is independent of the background field $\phi$ if $V^\prime(\phi \neq 0) = 0$ is renormalization scheme independent.  The way in which Stevenson's ``principle of minimal sensitivity'' (PMS) can be applied to the RG summed form of $R_{e{^+}e{^-}}$ is then discussed.  The significance of our results is considered in a concluding section. 
\end{abstract}

\section{Introduction}

Soon after it was established that divergences found present in quantum field theory calculations could be removed through the process of renormalization, it was realized that ambiguities arose at any finite order of perturbation theory.  Requiring that physical quantities be independent of parameters characterizing these ambiguities has led to the RG equations [1-3].  The parameter most usually considered is $\mu$, the mass scale introduced in the course of renormalization (for an interesting perspective on this see ref. [4]) but the additional ambiguities that arise in quantum chromodynamics (QCD), when using mass-independent renormalization [5,6], have been shown by Stevenson [7] to be parameterized by the coefficients $c_i(i \geq 2)$ of the expansion of the $\beta$-function associated with $\mu$ beyond two-loop order. The RG functions $\beta_i$ associated with these parameters $c_i$ can be expressed in terms of the $\beta$-function itself; certain linear combinations of renormalization scheme dependent parameters were shown to be renormalization scheme independent [7,8].

Various approaches have been considered to mitigate the dependence of physical quantities at finite order in perturbation theory on the parameters characterizing the renormalization scheme used [7, 9-20]. The total cross section for $e^+e^-$ annihilation into hadrons provides a convenient example for testing the efficacy of these approaches [7, 21-23].

It has been shown that variation of physical quantities with changes in the scale parameter $\mu$ is reduced by using the RG equation associated with $\mu$ to sum so-called leading-log (LL), next-to-leading-log (NLL) etc. corrections that arise at arbitrarily high order of perturbation theory.  This was originally suggested by Maxwell [24], and has been carried out in various physical processes [25,26], as well as in the effective action for instantons [27], thermal field theory [28], and the Coleman-Weinberg potential [29-36].  (This summation procedure has also been used to show that when using the MS renormalization scheme to relate the bare and renormalized coupling in dimensional regularization [38], the bare coupling vanishes rather than diverges when the dimensionality of space-time approaches four [39].)

Instead of using the RG equation to sum all LL, NLL etc. contributions to a physical process, it is also possible to perform a sum of all logarithmic contributions to a physical process, leading to an expression in which only the log-independent parts explicitly contribute, along with an auxiliary ``running coupling'' that contains all log-dependent contributions and whose behaviour is governed by the usual RG function $\beta$.  This summation has been useful in considering the effective action [27] as well as the Coleman-Weinberg potential [33-37].

In this paper, we will consider application of the RG equations associated with the parameters $c_i$ to the RG summed expression for $R_{e{^+}e{^-}}$, using the second approach to summation. The RG functions $\beta_i$ in this case depend not only on the couplant $a$ but also on the parameters $c_i$ themselves.  This prevents one from integrating these RG equations even formally; this is unlike the situation for the RG equation associated with the mass scale $\mu$ as the $\beta$-function associated with $\mu$ depends solely on the couplant $a$ and is independent of $\mu$ itself.  However, it is possible to determine how the log independent contributions to 
 $R_{e{^+}e{^-}}$ depend on $c_i$, which in turn fixes the dependence of the log dependent contribution to $R_{e{^+}e{^-}}$ on these parameters.  
 
In the course of determining how $R_{e{^+}e{^-}}$ depends on the parameters $c_i$, renormalization scheme invariants $\tau_i$ arise.  We show how these $\tau_i$ are related to the renormalization scheme invariants $\rho_i$ found by Stevenson [7,8].

We now consider two different choices of renormalization scheme; that is, we consider two different ways of selecting the parameters $c_i$.  First, we can eliminate all dependence of $R_{e{^+}e{^-}}$ on effects beyond two-loop order, save for the dependence of the running coupling on the renormalization scheme independent parameters $\tau_i$.  Secondly, we can set all $c_i$ equal to zero, allowing one to expand $R_{e{^+}e{^-}}$ in powers of the two-loop running coupling (which is given by the Lambert $W$-function [28, 40, 41]) with coefficients dependent solely on the $\tau_i$.  In both cases, $R_{e{^+}e{^-}}$ is independent of the scheme dependent parameters $c_i$.  Upon comparing these two ways of expanding $R_{e{^+}e{^-}}$ with a general expression relating the running coupling in two different renormalization schemes; we find consistency between these two renormalization schemes used to compute $R_{e{^+}e{^-}}$.

Next we show how within a given renormalization scheme, the running coupling at one mass scale can be expanded in powers of the running coupling at a different mass scale. Upon using this result in conjunction with the expansion of the RG summed form of $R_{e{^+}e{^-}}$ we show how the mass scale at each order of the running coupling can be chosen to absorb all renormalization scheme dependency of $R_{e{^+}e{^-}}$.

This is in keeping with the approach known as the ``principle of maximum conformality'' [13-20] (PMC), though this way of handling renormalization scheme ambiguities was originally applied to a perturbative expansion of $R_{e{^+}e{^-}}$ in which RG summation had not been used.

The summation of all logarithmic corrections to the Coleman-Weinberg effective potential $V$ by use of the RG equation has led to the interesting result that $V$ is independent of the constant background field $\phi$ when the condition $V^\prime (\phi) = 0$ at $\phi = v$ is applied, provided $v \neq 0$.  This has been demonstrated in a number of models [33-37]; here we consider the simplest of these (a massless $\phi^4$ model in four dimensions) and show that the resulting expression for $V(\phi)$ is renormalization scheme independent.

We then consider how Stevenson's PMS approach to choosing the parameters ($\mu$, $c_i$) that characterize a renormalization scheme can be applied to $R_{e{^+}e{^-}}$ after performing each of the two RG summations that have been considered.

\section{Renormalization Group Summation}

To illustrate how renormalization scheme dependence occurs after employing RG summation, we will consider the usual example of the total cross section for $e^+e^-$ annihilation into hadrons, ignoring the threshold effects of heavy quarks and complications due to gauge choice, after normalizing it by the cross section for $e^+e^-$ annihilating into $\mu^+\mu^-$.  If this quantity $R_{e{^+}e{^-}}$ is written as [7]
\begin{equation}
R_{e{^+}e{^-}} = \left(3 \sum_i q_i^2\right) (1 + R)
\end{equation}
then $R$ can be expanded in powers of the couplant $a$
\begin{equation}
R = \sum_{n=0}^\infty r_n a^{n+1} \quad (r_0 = 1)
\end{equation}
where $r_n$ contains the $n$ loop contribution to $R$.  By considering the Feynman diagrams that contribute to $R$ one can see that $r_n$ is given by 
\begin{equation}
r_n = \sum_{m=0}^n T_{nm} L^m
\end{equation}
where $T_{00} = 1$ and $L \equiv b\log(\mu/Q)$ where $Q$ is the centre of mass energy for $R_{e{^+}e{^-}}$.  As $R$ is independent of the renormalization mass scale $\mu$, we have the RG equation
\begin{equation}
 \mu \frac{d}{d\mu}R  = \left( \frac{\partial}{\partial \mu} + \beta(a) \frac{\partial}{\partial a} \right) R = 0
\end{equation}
where
\begin{equation}
\beta (a) = \mu \frac{\partial a}{\partial \mu} = - b a^2 \left(1 + c a + c_2 a^2 + c_3 a^3 + \ldots \right).
\end{equation}
In ref. [7] it is shown that $c_2, c_3, \ldots$ characterize the renormalization scheme ambiguities that reside in $R$ when it is computed to finite order in perturbation theory when using mass independent renormalization.  To show that $b$ and $c$ in eq. (5) are renormalization scheme independent, one considers the couplings $a$ and $\overline{a}$ associated with two different schemes so that
\begin{equation}
\overline{a} = a + x_2 a^2 + x_3 a^3 + \ldots .
\end{equation}
If
\begin{subequations}
\begin{align}
\overline{\beta}(\overline{a}) = \mu \frac{\partial \overline{a}}{\partial\mu}\\
\intertext{then together eqs. (5, 6, 7a) show that}
\overline{\beta}(\overline{a}) &= - \overline{b} \overline{a}^2 ( 1 + \overline{c}\;
\overline{a} + \overline{c}_2\overline{a}^2 + \ldots )\nonumber \\
&= (1 +2x_2 a + 3x_3 a^3 + \ldots ) (-ba^2) (1 + ca+c_2 a^2 + \cdots)
 \end{align}
\end{subequations}
which are compatible if $b = \overline{b}$, $c = \overline{c}$ while $c_2 = \overline{c}_2 + cx_2 + x^2_2 - x_3$ etc.

\subsection{Relations between $T_{nm}$}

We now will consider eq. (4) in more detail, and show how it can be used to fix $T_{nm}(1 \leq m \leq n)$ in terms of $T_{n0}$.  We can write eq. (4) as 
\begin{align}
\left( \mu \frac{\partial}{\partial \mu}  - ba^2(1 + ca + c_2 a^2 + \ldots )\frac{\partial}{\partial a}\right) \sum_{n=0}^\infty \sum_{m=0}^n a^{n+1} T_{nm}L^m\nonumber \\
 \qquad = \sum_{n=0}^\infty \sum_{m=0}^n \big[ m\,a^{n+1} T_{nm} L^{m-1} - a^2 (1 + ca + c_2a^2 + \ldots )\nonumber \\
 \left( (n+1) T_{nm} a^n L^m \right)\big] = 0.
\end{align}
By considering individual terms of order $a^pL^q$ in eq. (8), relations such as 
\begin{subequations}
\begin{align}
T_{ii} &= T_{i-1,i-1} \\
T_{21} &= (c + 2 T_{10})\\
2T_{32}& = (2cT_{11} + 3T_{21})\\
\intertext{and}
T_{31} &= c_2 + 3T_{20} + 2cT_{10}
\end{align}
\end{subequations}
follow. 
\subsection{Leading-log etc. Summation}

We now can systematically sum contributions to $R$ using the RG eq. of eq. (4).  We first define, as in ref. [25, 26, 37] functions 
\begin{equation}
S_n = \sum_{k=0}^\infty T_{n+k,k} (aL)^k
\end{equation}
so that eq. (2) becomes
\begin{equation}
R = R_\Sigma = \sum_{n=0}^\infty a^{n+1} S_n(aL)
\end{equation}
by eq. (3).  $S_0$ is the leading-log (LL) contribution to $R$, $S_1$ the next-to-leading-log (NLL) contribution to $R$, $S_p$ the N$^p$LL contribution etc.  Substitution of eq. (11) into eq. (4) leads to
\begin{subequations}
\begin{align}
&S_0^\prime - (S_0 + u S_0^\prime) = 0\\
&S_1^\prime - (2S_1 + u S_1^\prime) - c (S_0 +uS_0^\prime) = 0\\
&S_2^\prime - (3 S_2 + uS_2^\prime) - c(2S_1 + u S_1^\prime) - c_2 (S_0 + u S_0^\prime) = 0\\
& \mathrm{etc.}\nonumber
\end{align}
\end{subequations}
Solving these equations sequentially leads to [25, 26]
\begin{subequations}
\begin{align}
&S_0 = \frac{T_{00}}{w} \quad (w = 1 - u)\\
&S_1 = \frac{T_{10} - c T_{00} \ln |w|}{w^2}\\
&S_2 = \frac{T_{20} - (2c T_{10} + c^2 T_{00})\ln |w| + (c^2 - c_2) T_{00} (w-1) + c^2T_{00} \ln^2 |w|}{w^3}\\
& \mathrm{etc.}\nonumber
\end{align}
\end{subequations}
This shows how $T_{n+k,k}(k = 1, 2 \ldots)$ is determined by $T_{n0}$ in addition to $b, c, c_2 \ldots c_n$.  In ref. [26] it is demonstrated how
\begin{subequations}
\begin{align}
R^{[M]} = \sum_{n=0}^M a^{n+1} r_n\\
\intertext{varies more widely with changes in $\mu$ than does}\nonumber \\
R^{[M]}_{\sum} = \sum_{n=0}^M a^{n+1} S_n (aL).
\end{align}
\end{subequations}
This is to be expected as $R$ itself is independent of $\mu$, and so $R_{\sum}^{[M]}$ is necessarily a closer approximation to the exact expression for $R$ than $R^{[M]}$, containing as it does more terms in the expansions of eqs. (2,3).

\subsection{Summation of All Logarithms}

In place of the groupings of eq. (10), we consider a second grouping [27] 
\begin{equation}
A_n = \sum_{m=0}^\infty  T_{n+m,n} a^{n+m+1}
\end{equation}
so that
\begin{equation}
R = R_A = \sum_{n=0}^\infty A_n(a)L^n.
\end{equation}
Substitution of eq. (16) into eq. (4) now leads to 
\begin{equation}
\sum_{n=0}^\infty \left( b\,n A_n(a)L^{n-1} + \beta(a) A^\prime_n (a) L^n \right) = 0.
\end{equation}
This is satisfied at order $L^{n-1}$ provided
\begin{equation}
A_n(a) = - \frac{\beta(a)}{bn} \frac{d}{da} A_{ n-1}(a) .
\end{equation}
If now a parameter $\eta$ is introduced,
\begin{equation}
\eta \equiv \int_{a_{I}}^{a(\eta)} \frac{dx}{\beta(x)} \quad (a_I = \mathrm{const.})
\end{equation}
then
\begin{equation}
\beta(a) \frac{d}{da} = \frac{d}{d\eta}
\end{equation}
and so eq. (18) becomes
\begin{equation}
A_n(a) = \frac{-1}{bn}\frac{d}{d\eta} A_{n-1} (a(\eta)) = \frac{1}{n!}\left( - \frac{1}{b}\frac{d}{d\eta}\right)^n A_0 (a(\eta)).
\end{equation}
From eq. (16) then
\begin{equation}
 R_A  = \sum_{n=0}^\infty \frac{1}{n!} \left( - \frac{L}{b}\right)^n \frac{d^n}{d\eta^n} A_0 (a(\eta))
\end{equation}
\begin{equation}
= A_0 \left( a\left( \eta - \frac{1}{b} L\right)\right).
\end{equation}
This further demonstrates how $R$ depends on $T_{nm} (1 \leq m \leq n)$ only indirectly as $T_{nm} (1 \leq m \leq n)$ is fixed in terms of $T_{n0}$.  We note that dependence of $R$ on $\eta$ in eq. (23) can be absorbed into dependence on $\mu$ as
\begin{equation}
\eta - \frac{1}{b} L = -\frac{1}{b} L^\prime \equiv -\frac{1}{b} \log \left( \frac{\mu^\prime}{Q}\right)
\end{equation}
where $\mu^\prime =e^{ -\eta b}\mu$.

The function $a(\eta)$ introduced in eq. (19) satisfies
\begin{equation}
\frac{d a(\eta)}{d\eta} = \beta (a (\eta))
\end{equation}
but it is distinct from the ``running couplant'' $a$ originally appearing in eq. (2).  The $a$ in eq. (2), as it also appears in the RG equation (4), satisfies
\begin{equation}
\mu \frac{da(\mu)}{d\mu} = \beta (a(\mu))
\end{equation}
and has a boundary condition [7, 4] that involves a scale parameter $\Lambda$.  It is suggested in ref. [7] that the solution to eq. (26) is taken to be 
\begin{equation}
\ln \left( \frac{\mu}{\Lambda}\right) = \int_0^a \frac{dx}{\beta(x)} + \int_0^\infty \frac{dx}{bx^2(1+c x)},
\end{equation}
which is convergent for $a \neq 0$.  It is apparent that the $a(\eta)$ appearing in eqs. (19, 25) is distinct from $a(\mu)$ appearing in eqs. (26,27) even though they satisfy differential equations that have the same form. To distinguish the two, we will henceforth denote $a(\eta)$ appearing in eqs. (19, 25) by $\alpha(\eta)$ so that eq. (23) becomes
\begin{equation}
R = R_A = A_0 \left( \alpha \left( - \frac{1}{b}L^\prime\right)\right).
\end{equation}
The $a_I$ appearing in eq. (19) as an integration constant to the differential equation of eq. (25) can be seen by setting $Q = \mu^\prime$ in eqs. (2, 3, 28) to be just function $a(\mu)$ appearing in eqs. (26, 27).  Thus when we use $R_A$ to compute the $e^+e^-$ annihilation cross section, the value of $\Lambda$ is not in itself relevant; rather we should concern ourselves with the value of $a(\mu)$ which is the boundary value of $\alpha$ when $Q = \mu^\prime$ (using the renormalization scheme parameterized by $c_i$).

From now on we will drop the prime on $\mu^\prime$ and $L^\prime$.

\section{Renormalization Scheme Dependence}

As has been shown above, in the expansion of $\beta(\alpha)$ given by eq. (5), $b$ and $c$ are independent of the renormalization scheme used; in ref. [7] Stevenson has demonstrated that the constants $c_i$ provide a complete set of parameters characterizing any renormalization scheme in massless QCD provided it is a mass independent scheme.  We will now show explicitly how $A_0\left(\alpha (-\frac{1}{b} L)\right)$ in eq. (28) depends on $c_i$.

If the RG function $\beta_i(\alpha , c_i)$ is defined by
\begin{equation}
\beta_i (\alpha, c_j) = \frac{\partial \alpha}{\partial c_i}
\end{equation}
then as
\begin{equation}
\left( \frac{\partial^2}{\partial \eta\partial c_i} - \frac{\partial^2}{\partial c_i \partial\eta}\right) \alpha = 0
\end{equation}
it follows from eqs. (25, 29) that [7, 42]
\begin{align}
\beta_i (\alpha, c_j) &= -b \beta(\alpha) \int_0^\alpha \frac{dx\, x^{i+2}}{\beta^2(x)}\\
&= \frac{\alpha^{i+1}}{i-1} \sum_{n=0}^\infty W_n^i \alpha^n
\end{align}
where $W_0^i = 1$ and 
\begin{equation}
W_j^i = \left\vert
\begin{array}{cccl}
-(-2|0)c & +(-3|1)c_2 & -(-4|2)c_3\ldots & (-1)^j(-j-1|j-1)c_j \\
1 & -(-1|1)c & +(-2|2)c_2 \ldots &(-1)^{j+1} (-j+1|j-1)c_{j-1} \\
 & 1 & -(0|2)c \ldots &(-1)^j (-j+3|j-1)c_{j-2}\\
\dots & \ldots & \ldots & \ldots \\
 & & 1 & -(j-3|j-1) c
 \end{array}
 \right\vert
\end{equation}
with 
\begin{equation}
(m|n) \equiv (i + m)/(i+n).
\end{equation}
The first few terms contributing to eq. (32) are
\begin{align}
 \beta_j (\alpha , c_i)& = \frac{\alpha^{j+1}}{j-1}\bigg[ 1 + \frac{(-j+2)c}{j} \alpha + \frac{(j^2-3j+2)c^2+(-j^2+3j)c_2}{(j+1)j} \alpha^2 \\
&+ \frac{c_3(-j^3 + 3j^2 +4j)+cc_2(2j^3-6j^2+4)+c^3(-j^3+3j^2-2j)}{(j+2)(j+1)j} \alpha^3  + \ldots \bigg]. \nonumber
\end{align}

We now have the requirement that
\begin{equation}
\frac{d}{dc_i} R = \left( \frac{\partial}{\partial c_i} + \beta_i (\alpha , c_j)\frac{\partial}{\partial\alpha}\right) R_A = 0
\end{equation}
which by eqs. (15, 28, 32) becomes (with $T_n \equiv T_{n0}$)
\begin{equation}
\left( \frac{\partial}{\partial c_i} + \frac{\alpha^{i+1}}{i-1} \sum_{j=0}^\infty  W_j^i \alpha^j \frac{\partial}{\partial \alpha}\right)
\left[ \sum_{n=0}^\infty T_n \alpha^{n+1}   \right] = 0.
\end{equation}
By considering terms of order $\alpha^{i+j}$ in eq. (37) we find that
\begin{align}
\frac{\partial T_{i+j}}{\partial c_i} + \frac{1}{i-1} \big[ (1) W_j^i T_0 + (2) W_{j-1}^i T_1 + (3) W_{j-2}^i T_2\\
+ \ldots +(j+1) W_0^i T_j \big] = 0\nonumber
\end{align}
with $W_j^i$ given by eq. (33).

From eq. (38), we find that
\begin{equation}
\frac{\partial T_2}{\partial c_2} + 1 = 0
\end{equation}
which shows that
\begin{equation}
T_2 = -c_2 + \tau_2
\end{equation}
where $\tau_2$ is a constant of integration for eq. (39).
It then follows from eq. (38) that
\begin{subequations}
\begin{align}
\frac{\partial T_3}{\partial c_2} + 2\tau_1 = 0 \\
\intertext{and}
\frac{\partial T_3}{\partial c_3} + \frac{1}{2} = 0 
\end{align}
\end{subequations}
and so
\begin{equation}
T_3 = -2c_2 \tau_1 -\frac{1}{2} c_3 + \tau_3 ;
\end{equation}
similarly
\begin{subequations}
\begin{align}
&\frac{\partial T_4}{\partial c_2}+ \frac{1}{3} c_2 + 3T_2 = 0 \\
&\frac{\partial T_4}{\partial c_3} + \frac{1}{2}\left(- \frac{1}{3} c + 2 T_1\right) = 0 \\
&\frac{\partial T_4}{\partial c_4} + \frac{1}{3}  = 0 
\end{align}
\end{subequations}
show that
\begin{equation}
T_4 = - \frac{1}{3} c_4 -\frac{c_3}{2}\left(-\frac{1}{3} c + 2 \tau_1\right) +
 \frac{4}{3} c_2^2 - 3 c_2 \tau_2 + \tau_4
\end{equation}
and
\begin{subequations}
\begin{align}
&\frac{\partial T_5}{\partial c_2} + \left( -\frac{1}{6} c_2 c + \frac{1}{2} c_3\right) + 2 \left(+ \frac{1}{3} c_2 \right) T_1 + 4 T_3 = 0\\
&\frac{\partial T_5}{\partial c_3} +  \frac{1}{2} \left[  \frac{1}{6} c^2 + 2 \left(+ \frac{1}{3} c\right) T_1 + 3 T_2\right] = 0\\
&\frac{\partial T_5}{\partial c_4} +  \frac{1}{3}\left[ \left( - \frac{1}{2} c\right) +  2 T_1 \right]= 0\\
&\frac{\partial T_5}{\partial c_5} +  \frac{1}{4}  = 0
\end{align}
\end{subequations}
lead to
\begin{align}
T_5 &= \left[ \frac{1}{3} c c_2^2  + \frac{3}{2} c_2 c_3 + \frac{11}{3} c_2^2 \tau_1 - 4 c_2 \tau_3  \right]\nonumber \\
&- \frac{1}{2} \left[ \frac{1}{6}  c^2c_3 - \frac{2}{3} c_3 c \tau_1 + 3 c_3 \tau_2  \right]\nonumber \\
&-\frac{1}{3} \left[ -\frac{1}{2}  c_4 c + \frac{1}{2} c_4 \tau_1\right] - \frac{1}{4} c_5  +  \tau_5 . 
\end{align}
In eqs. (40, 42, 44, 46) $\tau_i$ are all constants of integration associated with the differential equations for $T_i$; they are renormalization scheme invariants as they are independent of $\mu$ and $c_i$.  To evaluate them, one must compute the Feynman diagrams associated with $R$ to the appropriate order in perturbation theory using the same renormalization scheme that has been used to determine the $c_i$; knowing $T_i$ and $c_i$ one can then solve for the $\tau_i$.

It is of interest to see how the renormalization invariants $\rho_i$ [7,8] are related to the $\tau_i$.  We first consider the invariant
\begin{align}
\rho_2 &= L - r_1\nonumber \\
& = T_{00} L - (T_{10} + T_{11} L).
\end{align}
The term in eq. (35) dependent on $L$ is satisfied by virtue of eq. (9a); the term independent of $L$ results in 
\begin{equation}
\rho_2 = -\tau_1.
\end{equation}
  Next, the invariant $\rho_3$ is given by 
\begin{equation}
\rho_3 = c_3 + 2r_3 -2c_2 r_1 - 6r_2r_1 + c r_1^2 + 4 r_1^3
\end{equation}
which, using eq. (3) becomes
\begin{align}
\rho_3 &= c_3 + 2 (T_{30} + T_{31} L + T_{32} L^2 + T_{33} L^3) - 2 c_2 (T_{10} + T_{11} L)\nonumber \\
& \quad -6 (T_{20} + T_{21} L + T_{22} L^2) (T_{10} + T_{11} L)\nonumber \\
&\qquad + c (T_{10} + T_{11} L)^2 + 4 (T_{10} + T_{11} L)^3.
\end{align}
From eqs. (9a-d) one finds that eq. (50) is satisfied due to eq. (4) at orders $L$, $L^2$ and $L^3$; from the terms independent of $L$ we find from eq. (50) that $\rho_3$ can be expressed in terms of $\tau_1$, $\tau_2$, $\tau_3$, and $c$ using eqs. (40, 42).  It is independent of $c_2$ and $c_3$, as it should, as these parameters are scheme dependent. This pattern should persist for all $\rho_n$.

We now consider two special values for the parameters $c_i$ which characterize our choice of renormalization scheme.  First of all, the $c_i$ can be expressed in terms of the $\tau_j$ so that $T_n = 0$ for all $n \geq 2$.  From eqs. (40, 42, 44, 46) this means that
\begin{subequations}
\begin{align}
c_2 &= \tau_2\\
c_3 &= 2(-2 c_2 \tau_1 + \tau_3)\nonumber \\
&= -4 \tau_2\tau_1 + 2\tau_3\\
c_4 &= 3 \bigg[ - \frac{c_3}{2}\left( - \frac{1}{3} c + 2 \tau_1 \right) + \frac{4}{3} c_2^2 \nonumber\\
&\quad- 3 c_2\tau_2 + \tau_4 \bigg]\nonumber \\
&\quad= c(\tau_3 - 2 \tau_1\tau_2) + 12 \tau_1^2 \tau_2 - 6 \tau_1\tau_3 - 5 \tau_2^2 + 3\tau_4 \\
\intertext{and}
c_5 &= 4 \bigg\{ \left[ \frac{1}{3} cc_2^2 + \frac{3}{2} c_2c_3 + \frac{11}{3}c_2^2 \tau_1 - 4c_2\tau_3 \right]\nonumber \\
&\qquad-\frac{1}{2}\left[ \frac{1}{6} c^2c_3 - \frac{2}{3}c_3 c  \tau_1 +3 c_3\tau_2 \right]\nonumber \\
&\qquad-\frac{1}{3}\left[- \frac{1}{2} c_4c + \frac{1}{2}c_4   \tau_1\right] +\tau_5\bigg\}\nonumber\\
&= \left[ \frac{4}{3} c\tau_2^2 + \frac{44}{3} \tau_2^2\tau_1 - 16 \tau_2\tau_3 \right]\nonumber \\
&\quad+ \left[2\tau_3- 4 \tau_1\tau_2 \right]
 \left[ 6\tau_2 - \frac{1}{3} c^2 + \frac{4}{3} c\tau_1 - 6 \tau_2 \right]\nonumber\\
&\quad +\left[c(\tau_3 - 2 \tau_1\tau_2) + 12 \tau_1^2\tau_2 - 6 \tau_1\tau_3 - 5 \tau_2^2 + 3\tau_4 \right]\left[ \frac{2}{3} (c - \tau_1)\right] + 4\tau_5
 \end{align}
 \end{subequations}
with $c_6$, $c_7$ etc. being computed in a similar fashion.  This leads to complicated expressions for the $c_i$ in terms of the renormalization scheme invariants $\tau_i$, but the full expression for $R$ collapses down to just two terms
\begin{equation}
R = R_A^{(1)} = \alpha_{(1)} \left( - \frac{1}{b} L\right) + \tau_1 \alpha_{(1)}^2 \left( - \frac{1}{b} L\right)
\end{equation}
by eq. (28).  ($\alpha_{(1)}$ denotes the running couplant $\alpha$ with this first choice of $c_i$.)

A second choice for the $c_i$ is to simply set $c_i = 0$ [43].  In this case, $T_n = \tau_n$ and we have the running couplant given exactly by (from eq. (19))
\begin{equation}
\eta = \int_a^{\alpha_{(2)}(\eta)} \frac{dx}{-bx^2(1+c x)} ,
\end{equation}
from which we obtain [28, 40, 41]
\begin{equation}
\left(-1-\frac{1}{c\alpha_{(2)}}\right) e^{-1-\frac{1}{c\alpha_{(2)}}} = e^{-\frac{b\eta}{c}} \left(-1-\frac{1}{ca}\right)e^{-1-\frac{1}{ca}}
\end{equation}
showing that with this second choice for $c_i$, the running coupling $\alpha_{(2)}(\eta)$ can be expressed in  terms of the Lambert function $W(x)$ (i.e., $x = W(x)e^{W(x)}$ [41]).  With all $c_i = 0$, then $T_i = \tau_i$ and so now
\begin{equation}
R = R_A^{(2)} = \alpha_{(2)} + \tau_1 \alpha_{(2)}^2 +  \tau_2 \alpha_{(2)}^3 +  \tau_3 \alpha_{(2)}^4 + \ldots\, .
\end{equation}
Unlike the expression for $R_A^{(1)}$ given in eq. (52), this expression for $R_A^{(2)}$ involves an infinite series, though $R_A^{(2)}$ does have the advantage that the couplant $\alpha_{(2)}$ is known exactly.  Both $R_A^{(1)}$ and $R_A^{(2)}$ depend exclusively on the renormalization scheme invariants $b$, $c$, $\tau_i$ which are to be determined through the evaluation of Feynman diagrams.

We now will derive a general relationship between couplants $\alpha_{(c)}$ and $\alpha_{(d)}$ evaluated using different renormalization schemes characterized by the parameters $c_i$ and $d_i$ respectively.  To do this, we make an expansion
\begin{equation}
\alpha_{(c)} = \alpha_{(\alpha)} + \lambda_2 (c_i, d_i) \alpha^2_{(d)} + \lambda_3 (c_i, d_i) \alpha_{(d)}^3  + \ldots
\end{equation}
Since $\alpha_{(c)}$ is independent of $d_j$, then
\begin{equation}
\frac{d}{dd_j} \alpha_{(c)} = \left( \frac{\partial}{\partial d_j} + \beta_j (d_i)  
\frac{\partial}{\partial\alpha_{(d)}} \right) \sum_{N=1}^\infty \lambda_N (c_i, d_i) \alpha^N_{(d)} = 0 
\end{equation}
where $\beta_j(d_i)$ is given by eqs. (32-35).  Eqs. (35) and (57) together show that
\[
 \frac{\partial\lambda_2}{\partial d_2} = 0, \quad 
\frac{\partial\lambda_3}{\partial d_2} + 1 = 0, \quad 
\frac{\partial\lambda_4}{\partial d_2} + 2\lambda_2 = 0, \quad 
\frac{\partial\lambda_5}{\partial d_2} + \frac{d_2}{3} +3\lambda_3 = 0 \eqno(58a-d)
\]
\[ \frac{\partial\lambda_2}{\partial d_3} = 0, \quad 
\frac{\partial\lambda_3}{\partial d_3}  = 0, \quad 
\frac{\partial\lambda_4}{\partial d_3} + \frac{1}{2} = 0, \quad 
\frac{\partial\lambda_5}{\partial d_3} + \frac{1}{2}\left( - \frac{c}{3} + 2\lambda_2 \right) = 0 \eqno(59a-d)\]
\[\hspace{-5cm} \frac{\partial\lambda_2}{\partial d_4} = 0, \quad 
\frac{\partial\lambda_3}{\partial d_4}  = 0, \quad 
\frac{\partial\lambda_4}{\partial d_4}  = 0, \quad 
\frac{\partial\lambda_5}{\partial d_4} + \frac{1}{3} = 0 \eqno(60a-d)\]
etc.  (We see that $\lambda_{N-1}$ can depend on $d_2 \ldots d_N$.)  Solving eqs. (58-60) subject to the boundary condition
\[ \lambda_N (c_i,c_i) = 0 \eqno(61) \]
leads to 
\[ \alpha_{(c)} = \alpha_{(d)} - (d_2 - c_2) a^3_{(d)} - \frac{1}{2} (d_3 - c_3) \alpha^4_{(d)}\nonumber \]
\[+ \bigg[ - \frac{1}{6} \left( d_2^2 - c_2^2\right) + \frac{3}{2} \left(d_2 - c_2\right)^2 + \frac{c}{6} \left(d_3 - c_3 \right)\nonumber \]
\[ - \frac{1}{3} \left(d_4 - c_4 \right) \bigg] \alpha^5_{(d)} + \ldots . \eqno(62) \]
If eq. (62) is used to expand $\alpha_{(d)}$ in terms of $\alpha_{(e)}$ and then $\alpha_{(d)}$ is eliminated in eq. (62), then the resulting expansion of $\alpha_{(c)}$ in terms of $\alpha_{(e)}$ is also of the form of eq. (62).  This is a useful consistency check.

If in eq. (62) we were to $d_i = 0$ and chose $c_i$ so that $T_n = 0\;\, (n = 2,3 \ldots)$ (i.e., $c_2 \ldots c_5$ are given by eq. (31)) then we have an expansion for $\alpha_{(1)}$ in terms of $\alpha_{(2)}$.  but since the expansions for $R$ given in eqs. (52) and (55) can be equated, we also have 
\[ \alpha_{(1)} + \tau_1 \alpha_{(1)}^2 = \alpha_{(2)} + \tau_1 \alpha_{(2)}^2 + \tau_2 \alpha_{(2)}^3 + \tau_3 \alpha_{(2)}^4 + \ldots \eqno(63) \]
It can be shown that eqs. (62) and (63) are compatible upon identifying $\alpha_{(c)}$ and $\alpha_{(d)}$ in eq. (62) with $\alpha_{(1)}$ and $\alpha_{(2)}$ respectively in eq. (63), demonstrating the two renormalization schemes used to compute $R$ are consistent.

\section{Varying of Mass Scales and RG Summation}

It has been suggested that in the standard perturbative expansion, such as the one for $R$ in eq. (2), the mass scale $\mu$ chosen at each order of perturbation theory could be different and that by an appropriate selection of mass scales, all dependency on the renormalization scheme parameters can be absorbed into these mass scales.  We now will examine how this approach can be applied to the RG summed form of $R$ given by $R_A$ in eq. (28).

To begin with, we note how $\alpha_0 \equiv \alpha \left( - \frac{1}{b} \log \frac{\mu}{Q}\right)$ appearing in eq. (28) can be expanded in terms of  
$\alpha_i \equiv \alpha \left( - \frac{1}{b} \log \frac{\nu_i}{Q}\right)$ associated with the mass scale $\nu_i$ in the following way [27, 18, 44]
\[ \alpha_0 = \alpha_i + \left(\sigma_{21} \ell_{0i} \right) \alpha_i^2 + \left(\sigma_{31} \ell_{0i} + \sigma_{32} \ell_{0i}^2 \right)\alpha_i^3\eqno(64)\]
\[ + \left(\sigma_{41} \ell_{0i} + \sigma_{42} \ell_{0i}^2 +\sigma_{43} \ell_{0i}^3 \right) \alpha_i^4 + \ldots \nonumber \]
where $ \ell_{0i} \equiv \ln \left( \frac{\nu_i}{\mu}\right)$.  The coefficients $\sigma_{mn}$ can be fixed by noting that $\alpha_0$ is independent of $\nu_i$ so that
\[ \nu_i \frac{d}{d\nu_i} \alpha_0 = \left[ \nu_i \frac{\partial}{\partial \nu_i}- \frac{1}{b} \beta (\alpha_i) \frac{\partial}{\partial \alpha_i} \right] \sum^\infty_{m=1} \sum_{n=0}^{m-1} \sigma_{mn} \ell_{0i}^n a^m_i \quad \left(\sigma_{m0} = \delta_{m0}\right) =0. \eqno(65)\]
From eq. (65) we find that
\[ \alpha_0 = \alpha_i - \ell_{0i} \alpha_i^2 + \left( -c \;\ell_{0i} + \ell_{0i}^2 \right) \alpha_i^3 + 
\left( -c_2\ell_{0i} + \frac{5}{2} c\; \ell_{0i}^2 - \ell_{0i}^3 \right) \alpha_i^4\nonumber \]
\[+ \left( -c_3 \ell_{0i} + \left( 3c_2 + \frac{3}{2} c^2\right) \ell_{0i}^2 - \frac{13}{3} c \;\ell_{0i}^3 + \ell_{0i}^4\right) \alpha_i^5 \nonumber \]
\[+ \bigg( -c_4 \ell_{0i} + \frac{7}{2}\left( c_3 +c_2c\right)\ell_{0i}^2 - \left(6c_2 + \frac{35}{6} c^2\right) \ell_{0i}^3 \nonumber \]
\[+ \frac{77}{12} c \;\ell_{0i}^4 - \ell_{0i}^5\bigg) \alpha_i^6 + \ldots \;. \eqno(66)\]

Similarly , $\alpha_i$ can be expanded in terms of $\alpha_j$ with $\ell_{0i}$ in eq. (66) being replaced by $\ell_{ij} = \ln \left(\frac{\nu_j}{\nu_i}\right)$.  If eq. (66) is used to expand $\alpha_0$ in terms of $\alpha_i$ and then into this expansion we substitute the expansion of $\alpha_i$ in terms of $\alpha_j$, we obtain an expansion of $\alpha_0$ in terms of $\alpha_j$ which has the form of eq. (66) upon using $\ell_{ij} = \ell_{0j} - \ell_{0i}$ (as expected).

We also note that summations of terms in eq. (66) similar to those in eqs. (11) and (23) is possible [27].

The expansion of $R$ in eq. (28), upon using eqs. (40,42,44,46) can be written as
\[\hspace{-1cm} R = \alpha_0 + \tau_1 \alpha_0^2 + (-c_2 + \tau_2) \alpha_0^3 + \left( -2c_2 \tau_1 - \frac{1}{2} c_3 + \tau_3\right) \alpha_0^4\eqno(67)\]
\[ \qquad + \left[ -\frac{1}{3} c_4 - \frac{c_3}{2} \left(- \frac{1}{3} c + 2 \tau_1\right) + \frac{4}{3} c_2^2 - 3 c_2 \tau_2 +\tau_4\right] \alpha_0^5\nonumber \]
\[\hspace{2cm} + \bigg[ \left(\frac{1}{3} cc_2^2 + \frac{3}{2} c_2c_3 + \frac{11}{3} c_2^2 \tau_1 - 4 c_2\tau_3  \right) - \frac{1}{2} \bigg( \frac{1}{6} c^2 c_3 - \frac{2}{3} c_3 c \tau_1 \nonumber \]
\[\hspace{2cm}+ 3c_3\tau_2 \bigg) - \frac{1}{3} \left( - \frac{1}{2} c_4 c + \frac{1}{2} c_4\tau_1\right) - \frac{1}{4} c_5 + \tau_5 \bigg] \alpha_0^6 + \ldots \,.\nonumber \]

We now use eq. (66) to re-express $(\alpha_0)^N$ wherever it occurs in eq. (67) as $(\alpha_N )^N$.  For example, we can have 
\[ \alpha_0 = \alpha_1 - \ell_{01} \alpha_1^2 + (-c\;  \ell_{01} + \ell_{01}^2) \alpha_1^3 + (-c_2 \ell_{01} + \frac{5}{2} c \;\ell_{01}^2 - \ell_{01}^3) \alpha_1^4 + \ldots\nonumber \]
\[\hspace{-1.5cm}= \alpha_1 - \ell_{01} \left[ \alpha_2 - \ell_{12} \alpha_2^2 +( -c \;\ell_{12} + \ell_{12}^2) \alpha_2^3 + \ldots\right]^2\eqno(68) \]
\[\hspace{-1.5cm} + (-c \;\ell_{01} + \ell_{01}^2) \left[ \alpha_3 - \ell_{13} \alpha_3^2 + \ldots \right]^3\nonumber\]
\[ + \left( -c_2 \ell_{01} + \frac{5}{2} c \ell_{01}^2 - \ell_{01}^3 \right) \left[ \alpha_4 + \ldots \right]^4 + \ldots .\nonumber \]
Repeating this procedure we eventually find
\[\hspace{-1cm} \alpha_0 = \alpha_1 - \ell_{01} \alpha_2^2 + \left[2\ell_{01} \ell_{12} + (-c\; \ell_{01} + \ell_{01}^2)\right] \alpha_3^3 \eqno(69a) \]
\[+\bigg[ -6  \ell_{01}\ell_{12}\ell_{23}
 - \ell_{01} \left(  \ell_{12}^2 + 2 (-c \;\ell_{12} 
+ \ell_{12}^2)\right)\nonumber \]
\[\hspace{1cm} - 3 \ell_{13} ( - c  \;\ell_{01} +  \ell_{01}^2) + \left( -c_2 \ell_{01} + \frac{5}{2} c\; \ell_{01}^2 - \ell_{01}^3 \right)\bigg] \alpha_4^4 + \ldots \nonumber \]
as well as
\[ \alpha_0^2 = \alpha_2^2 - 2\ell_{02} \alpha_3^3 + \left[6\ell_{02} \ell_{23} +  \ell_{02}^2 + 
 2 (-c\; \ell_{02} + \ell_{02}^2)\right] \alpha_4^4 + \ldots \eqno(69b)\]
\[\hspace{-6.5cm} \alpha_0^3 = \alpha_3^3 - 3\ell_{03} \alpha_4^4 + \cdots \eqno(69c)\]
\[\hspace{-7.7cm} \alpha_0^4 = \alpha_4^4 + \ldots .\eqno(69d) \] 
(We keep only terms to order $\alpha^4$.)

Together, eqs. (67) and (69) result in
\[ R = \alpha_1 + \alpha_2^2 \left(- \ell_{01} + \tau_1\right) + \alpha_3^3  \bigg\lbrace 2\ell_{02} (\ell_{01} - \tau_1) - (c\; \ell_{01} + \ell_{01}^2)\eqno(70) \] 
\[+ (-c_2 + \tau_2)\bigg\rbrace + \alpha_4^4 \bigg\lbrace 3\ell_{03} \bigg[ 2\ell_{02} (-\ell_{01} + \tau_1) + \ell_{01}(\ell_{01}  + c)\nonumber \]
\[ - (-c_2 + \tau_2) \bigg] + \bigg[ 3\ell_{02}^2 (\ell_{01} - \tau_1) +2c \;\ell_{02} (\ell_{01} - \tau_1) - \ell_{01}^3 \nonumber \]
\[ - \frac{5}{2} c\;\ell_{01}^2 - c_2\ell_{01} + \left(-2 c_2 \tau_1 - \frac{1}{2} c_3 + \tau_3\right)\bigg]\bigg\rbrace + \ldots .\nonumber\]
(Again, we have used $\ell_{ij} = \ell_{0j} - \ell_{0i}$.)

In the PMC approach, the ambiguities in inherent in the parameters $c_i$ are absorbed into the mass scalars $\nu_i$.  It is easily seen how this can be done upon examining eq. (B.7).  At order $\alpha_1$ and $\alpha_2^2$, $R$ is independent of any $c_i$ and hence is unambiguous; however at order $\alpha_3^3$ the parameter $c_2$ explicitly occurs but it can be removed by choosing $\nu_2$ so that 
\[ 2 \ell_{02} (\ell_{01} - \tau_1) - (c \;\ell_{01} + \ell_{01}^2 ) + (-c_2) = 0 \eqno(71) \]
leaving the coefficient to $\alpha_3^3$ being the renormalization scheme invariant $\tau_2$.  Next, by choosing $\nu_3$ so that
\[\hspace{-2cm} 3\ell_{03} \left[ 2\ell_{02} (-\ell_{01} + \tau_1) + \ell_{01} (\ell_{01} + c) - (-c_2 + \tau_2)\right]\eqno(72) \]
\[ + \bigg[ 3\ell_{02}^2 (\ell_{01} - \tau_1) + 2c\;\ell_{02} (\ell_{01} - \tau_1) - \ell_{01}^3 - \frac{5}{2} c \;\ell_{01}^2 - c_2 \ell_{01}\nonumber \]
\[ + (-2 c_2 \tau_1 - \frac{1}{2} c_3) \bigg] = 0\nonumber \]
we eliminate all scheme dependence in the coefficient of $\alpha_4^4$ leaving solely the contribution coming from the renormalization scheme invariant quantity $\tau_3$.  This procedure can be applied at each order of $\alpha$ in the expansion of $R$; $\nu_j (j = 2,3 \ldots)$ can be selected to eliminate the dependence of the term of order $\alpha_{j+1}^{j+1}$ in the expansion of eq. (70) on $c_2, c_3 \ldots c_j$ leaving us solely with $\tau_j a_{j+1}^{j+1}$.  This should be possible for all $j$ as $\ell_{0j}$ enters the term of order $a_{j+1}^{j+1}$ only linearly.  We are left with
\[ R = \alpha_1 + (-\ell_{01} + \tau_1) \alpha_2^2 + \tau_2 \alpha_3^3 + \tau_4 \alpha_4^4 + \ldots \eqno(73) \]
which is reminiscent of eq. (55).  We note though that in eq. (73) the mass scales $\nu_j$ which enter $\alpha_j^j$ are scheme dependent as their value depends on the values of $c_2, c_3 \ldots c_j$.  In addition, $\alpha_1$, $\alpha_2$ etc. are necessarily dependent on the values of the $c_i$.  Thus unlike $R$ given by eq. (55), $R$ in eq. (73) retains an indirect renormalization scheme dependence.

The possibility of choosing $\nu_j$ so that the coefficient of $\alpha_{j+1}^{j+1}$ disappears completely for all $j$ should be considered.  However, this is not feasible (nor should we expect it to be, as $R$ would then reduce to being simply $\alpha_1$).  To see what happens if we attempt this, we note from eq. (70) that if we choose $\ell_{01}$ to eliminate the coefficient of $\alpha_2^2$, then the term of order  $\alpha_3^3$ loses its dependence on $\ell_{02}$.  Similarly, if $\ell_{02}$ is chosen to eliminate the term of order $\alpha_3^3$ (with $\ell_{01} \neq \tau_1$) then the coefficient of $\ell_{03}$ in the term of order $\alpha_4^4$ vanishes, making it impossible to select $\ell_{03}$ so that the term of order $\alpha_4^4$ vanishes.  This pattern should repeat itself at each term of order $\alpha_j^j$.

An interesting consistency check is to use eq. (62) to replace $\alpha\left( - \frac{1}{b} \log \frac{\nu_i}{Q}, c_j\right)$ by $\alpha\left( - \frac{1}{b} \log \frac{\nu_i}{Q}, d_j\right)$ in eq. (70); we find that eq. (70) is recovered with $d_j$ replacing $c_j$ and $\alpha\left( - \frac{1}{b} \log \frac{\nu_i}{Q}, d_j\right)$ replacing $\alpha\left( - \frac{1}{b} \log \frac{\nu_i}{Q}, c_j\right)$.

\section{Renormalization Scheme Ambiguities in the Effective Potential}

We will now examine renormalization scheme dependence in the Coleman-Weinberg effective action $V$ as considered in refs. [33-36]. Our attention will be restricted to a simple model in which a massless scalar field $\phi$ has a quartic self interaction so that the classical action is 
\[ S = \int d^4x  \left[\frac{1}{2}\left( \partial_\mu \phi \right)^2 - \frac{a}{4!} \phi^4\right].
\eqno(74) \]
The form that $V$ takes is
\[ V = \sum_{n=0}^\infty \sum_{m=0}^n T_{nm} a^{n+1} L^m \phi^4 \eqno(75) \]
with $L = \log \left(\frac{\phi}{\mu}\right)$ where now $\phi$ is the constant background field and $\mu$ is again the radiatively induced mass scale.  The $RG$ equation is
\[ \left( \mu \frac{\partial}{\partial \mu} + \beta(a) \frac{\partial}{\partial a} + \gamma (a) \frac{\partial}{\partial \phi} \right) V = 0\eqno(76) \]
when using a mass-independent renormalization scheme [5,6].  The $RG$ function $\beta (a)$ again has the form of eq. (5), while
\[ \gamma (a) = \frac{\mu}{\phi} \frac{\partial\phi}{\partial\mu} = fa \left(1 + g_1a + g_2a^2 + \ldots\right).\eqno(77) \]
Under the finite renormalizations of eq. (6) and
\[ \overline{\phi} = \phi\left(1 + y_1a + y_2a^2 + \ldots\right)\eqno(78) \]
it is apparent that  $b$, $c$, $f$ are unaltered while $c_2, c_3 \ldots, g_1, g_2 \ldots$ are all altered.  Following refs. [7, 42] (as well as ref. [45] for the case in which there is a mass to be renormalized) we characterize the renormalization scheme dependency by $c_2, c_3 \ldots$ and $g_1, g_2 \ldots$.  It is evident that $a$ is independent of $g_i$, while its dependency on $c_i$ is again given by eqs. (31-35); furthermore 
\[ \frac{\partial \phi}{\partial c_i} = \phi \gamma_i^c \eqno(79a)\]
and
\[ \frac{\partial \phi}{\partial g_i} = \phi \gamma_i^g .\eqno(79b)\]
Just as one can find $\beta_i$ from eq. (30), it follows from
\[ \left( \frac{\partial^2}{\partial\mu\partial c_i} -  
\frac{\partial^2}{\partial c_i\partial\mu}\right) \phi = \left( \frac{\partial^2}{\partial g_i\partial c_j} -  \frac{\partial^2}{\partial c_j\partial g_i}\right) \phi = \left( \frac{\partial^2}{\partial\mu\partial g_i} -  \frac{\partial^2}{\partial g_i\partial\mu}\right) \phi = 0 \eqno(80) \]
that
\[ \gamma_i^g = \int_0^a dx \frac{fx^{x+1}}{\beta(x)} \eqno(81a) \]
and
\[ \gamma_i^c = \frac{\gamma(a)\beta_i(a)}{\beta(c)} + b \int_0^a dx \frac{x^{i+2}\gamma(x)}{\beta^2(x)} .\eqno(81b)       \]

One can regroup the sum in eq. (75) as in eq. (10).  We will follow refs. [33, 37] and regroup the sum in eq. (75) as in eq. (15) so that
\[V = \sum_{n=0}^\infty A_n (a) L^n \phi^4 \eqno(82)    \]
with
\[ A_n(a) = \sum_{m=n}^\infty T_{mn} a^{m+1}.\eqno(83)\]
Eq. (36) now leads to 
\[ \hat{A}_{n+1} (a(\eta)) = \frac{1}{(n+1)} \frac{d}{d\eta} \hat{A}_n(a(\eta)) = \frac{1}{(n+1)!}\; \frac{d^{n+1}}{d\eta^{n+1}} \hat{A}_0(a(\eta))\eqno(84) \]
where
\[ \eta = \int_{a_{I}}^{a(\eta)} \frac{dx}{\hat{\beta}(x)} \qquad (a (\eta = 0) \equiv a_I)\eqno(85) \]
and
\[ \hat{A}_n (a) = A_n(a) \exp \left( 4 \int_{a_{I}}^a \frac{\hat{\gamma}(x)}{\hat{\beta}(x)} dx \right) \eqno(86) \]
where $\hat{\beta} = \beta / (1-\gamma)$ and $\hat{\gamma} = \gamma / (1-\gamma)$.  Together, eqs. (82) and (86) lead to 
\[ V = A_0 (a(\eta + L)) \exp \left( 4 \int_{a(\eta)}^{a(\eta + L)} 
\frac{\hat{\gamma}(x)}{\hat{\beta}(x)} dx \right) \phi^4 . \eqno(87) \]
By eq. (82), 
\[ \frac{dV}{d\phi} = \sum_{n=0}^\infty\left[ (n+1) A_{n+1} (a) + 4A_n(a)\right]L^n\phi^3; \eqno(88) \]
this vanishes at order $L^0$ when $\phi = v$ provided either $v = 0$ or
\[ A_1(a) + 4 A_0 (a)= 0.\eqno(89) \]

Together, eq. (84) when $n = 0$ and eq. (89) result in
\[ \left[ \hat{\beta}\frac{d}{da} + 4(1 + \hat{\gamma})\right] A_0(a) = 0 \eqno(90)\]
and 
\[ A_0 (a) = A_0 (a_I) \exp \left( -4 \int_{a_{I}}^a \frac{dx}{\beta(x)}\right).\eqno(91) \]
Together eqs. (87) and (91) result in 
\[V =  A_0 (a_I) \exp \left( -4 \int_{a_{I}}^a \frac{dx}{\beta(x)}\right) \mu^4 .\eqno(92) \]
Thus $V$ is independent of $\phi$ if $v \neq 0$.  It is immediately obvious that the $RG$ equation of eq. (76) is satisfied by $V$.

We now can examine the scheme dependence of eq. (92).  As $a$ is independent of $g_i$ and $V$ is independent of $\phi$, the equation
\[ \left( \frac{\partial}{\partial g_i} + \phi \gamma_i^g \frac{\partial}{\partial \phi} \right) V = 0 \eqno(93) \]
is automatically satisfied.  It is also clear that the equation
\[ \left( \frac{\partial}{\partial c_i} + \beta_i \frac{\partial}{\partial a} + \phi \gamma_i^c \frac{\partial}{\partial \phi} \right) V = 0 \eqno(94)\]
is satisfied on account of eq. (31).  Since eq. (92) satisfies eqs. (76), (93) and (94) it is wholly independent of all parameters that characterize the renormalization scheme being used.  We anticipate that this scheme independence also holds for the effective potential in more complicated models that have been considered such as a massive $\phi_4^4$ model [34-35], massless scalar electrodynamics [33] and the Standard Model with a single Higgs doublet of scalars [36].  In each of these models, $V$ has been shown to be independent of the background scalar field by using the appropriate versions of eqs. (76) and (88).

\section{The Principle of Minimal Sensitivity and RG Summation}

Stevenson in ref. [7] has proposed not only use of $(\mu , c_i)$ to characterize one's choice of renormalization  scheme in massless QCD when using mass independent renormalization, but also has argued that a ``principle of minimal sensitivity'' (PMS) be used to fix dependence of finite order perturbative results on these parameters.  When this approach has been applied to computations of $R_{e{^+}e{^-}}$ [21-23], the perturbative form of eq. (2) has been considered.  We will now see how PMS can be used with the RG summed form of $R$ given by eqs. (11) and (28) when only a finite number of terms contribute to these sums.

The cross section $R_{e{^+}e{^-}}$ has already been considered using the expansion of eq. (11) (as have the calculation of a number of physical quantities) [26].  It has been shown that variation of $R$ with changes of the scale parameter $\mu$ within a given renormalization scheme is considerably reduced when this RG summed form of $R$ is used instead of the perturbative result of eq. (2).  This is not unexpected, as the exact expression for $R$ must be independent of $\mu$ (and $c_i$), and since both RG summations include more contributions to $R$ than comes from an approximation arising from a truncated form of eq. (2), we should anticipate that the RG summed expressions have less dependency on $\mu$.

To consider this application of PMS more explicitly, let us examine the approximation 
\[ R_\Sigma^{(3)} = a S_0 (aL)+ a^2S_1 (aL) + a^3 S_2 (aL) \eqno(95) \]
to $R_\Sigma$ in eq. (11).  In more detail, by eqs. (13, 40) result in 
\[ R_\Sigma^{(3)} = \frac{a}{w} + a^2 \left(\frac{\tau_1 - c \ln |w|}{w^2}\right) \eqno(96) \]
\[ + a^3 \left( \frac{-c_2+\tau_2 - \left(2c\tau_1 + c^2\right) \ln |w| + \left(c^2-c_2\right) (w-1) + c^2 \ln^2 |w|}{w^3}\right) \nonumber \]
where $w = 1 - ab\log (\mu/Q)$ and to the order we are working by eqs. (27, 31)
\[ \ln \left( \frac{\mu}{\Lambda}\right) = \int_0^a \frac{dx}{-bx^2(1+cx+c_2x^2)} + \int_0^\infty \frac{dx}{bx^2(1+cx)} \eqno(97a) \]
and
\[ \frac{\partial a}{\partial c_2} = a^2 (1 + ca + c_2 a^2) \int_0^a \frac{dx}{(1+cx+c_2x^2)^2}. \eqno(97b) \]
In principle, the PMS criterion applied to $R_\Sigma^{(3)}$ involves applying the criterion
\[ \frac{\partial R_\Sigma^{(3)}}{\partial\mu}=  \frac{\partial R_\Sigma^{(3)}}{\partial c_2} = 0 \eqno(98)\]
in order to optimize the values of $\mu$ and $c_2$ in eq. (96).  Applying analyticly eq. (98) to $R_\Sigma^{(3)}$ in eq. (96) is clearly more difficult than applying the PMS criterion to the approximation
\[ R^{(3)} = a + r_1 a^2 + r_2 a^3 \eqno(99) \]
which follows from eq. (2) as was done in refs. [7, 21-23].

If in place of $R_\Sigma$ in eq. (11) one we to consider the RG sum of $R_A$ in eq. (28), then the approximation
\[ R_A^{(3)} = \sum_{n=0}^3 T_n \alpha^n \left( - \frac{1}{b} L\right)\eqno(100) \]
is of the same order as $R_\Sigma^{(3)}$ in eq. (95).  However, $R_\Sigma^{(3)}$ and $R_A^{(3)}$ are distinct quantities, having been derived using different RG summations.  More explicitly, using eqs. (40, 19, 31) we obtain
\[ R_A^{(3)} = \alpha \left( \ln \frac{Q}{\mu}\right) + \tau_1 \alpha^2 \left( \ln \frac{Q}{\mu}\right) + (-c_2 + \tau_2)\alpha^3 \left( \ln \frac{Q}{\mu}\right)\eqno(101) \]
where
\[  \ln\left( \frac{Q}{\mu}\right) = \int_a^\alpha \frac{dx}{-bx^2(1 + cx + c_2x^2)}, \eqno(102) \]
$\frac{\partial\alpha}{\partial c_2}$ is given by eq. (97b) with $a$ being replaced by $\alpha$, and the dependence of $a$ in eq. (102) on $\mu$ and $c_2$ itself being subject to eqs. (97a,b).

As with  $R_\Sigma^{(3)}$, applying the PMS criterion
\[ \frac{\partial R_A^{(3)}}{\partial \mu} = \frac{\partial R_A^{(3)}}{\partial c_2} = 0 \eqno(103) \]
analyticly is more involved than applying it to $R^{(3)}$ in (99).

\section{Discussion}

In this paper we have outlined two ways of performing RG summation of logarithmic contributions to the cross section $R_{e{^+}e{^-}}$; the final result in both cases involves simply the log-independent contribution to $R_{e{^+}e{^-}}$ and the RG function $\beta$.  Even though portions of $R_{e{^+}e{^-}}$ to arbitrarily high order in perturbation theory are incorporated by using these RG sums, the final result in both cases is not exact and consequently has explicit dependence on the parameters $\mu$ and $c_i$ that characterize the renormalization scheme used.  We have shown how these RG summed expressions for $R_{e{^+}e{^-}}$ depend on these parameters to any finite order in perturbation theory.  (We have also shown that the exact expression for $V$ in eq. (92) is renormalization scheme independent.)  In principle the PMS criterion can be applied to select ``optimal'' values of $\mu$ and $c_i$ but to do this would be non-trivial.

Of special interest are the two choices of the parameters $c_i$ that lead to the expansion $R_A^{(1)}$ and $R_A^{(2)}$ of eqs. (52) and (55) for $R$.  In both of these expansions, only the renormalization scheme invariants $b$, $c$ and $\tau_i$ appear; there is no dependency on $c_i$ either implicit or explicit.  All dependency on the physical moment $Q$ is in the argument of the auxiliary function $\alpha\left(\log \frac{Q}{\mu}\right)$; this function arises in the course of summing all of the logarithms appearing in eqs. (2) and (3).  The mass scale parameter $\mu$ only explicitly occurs in the ratio $Q/\mu$ and hence only serves to calibrate the magnitude of $Q$.  The couplant $a$ appearing in the expansion of eq. (2) is a boundary value for $\alpha(\eta)$ in eq. (25) and so all that is needed when considering $R_A^{(1)}$ and $R_A^{(2)}$ is the value of $a$ for some choice of $\mu$.  It is not necessary to consider how $a$ depends on $\mu$ through eq. (27) and the value of $\Lambda$ only reflects the value of $a$ at the value of $\mu$ chosen to calibrate the magnitude of $Q$.

The choice of $c_i$ that leads to $R_A^{(1)}$ in eq. (52) is appealing as $R$ then involves merely two terms; all $Q$ dependence of $R$ resides in the argument of $\alpha_{(1)}(\log Q/\mu)$ and its square.  This function has its behaviour dictated by the relation between the expansion constant $c_i$ and the renormalization scheme invariants $\tau_i$ (typified by eqs. (51a-d)).

A second choice for the $c_i$ is $c_i = 0$ [43] (though the feasibility of making this choice has been questioned in ref. [46]).  The auxiliary function $\alpha_{(2)} (\log Q/\mu)$ is now given in closed form (see. eq. (54)); the $\tau_i$ are now the expansion coefficients in the infinite series of eq. (55).  The simplicity of this result is again quite appealing.

We note that upon setting $Q = \mu$, so that by eq. (19)
\[ \alpha (Q = \mu) = a_I \eqno(104) \]
it follows by eqs. (52,55)
\[ R_\Sigma^{(1)} (Q = \mu) = a_{(1)} + \tau_1 a_{(1)}^2 \eqno(105a)\]
and
\[ R_I^{(2)} (Q = \mu) = a_{(2)} + \tau_1 a_{(2)}^2 + \tau_2 a_{(2)}^3 + \ldots\;. \eqno(105b) \]
Thus it is quite straight forward to determine $a_{(1)}$ (the value of $\alpha_{(1)}$ when $Q = \mu$) as it involves solving the quadratic in eq. (105a).  The value of $a_{(2)}$ is determined from eq. (105b); $a_{(1)}$ and $a_{(2)}$ are related due to eq. (63).

Clearly much can now be done. The ideas presented should be applied to a full quantitative analysis of the cross section $R_{e{^+}e{^-}}$. It would also be of interest to extend this approach to renormalization scheme dependency to processes in which there are non-trivial masses and/or multiple couplings in the classical action.  The scheme dependency occurring in the $\beta$-function of $N = 1$ supersymmetry [47, 48] and the effective action for an external gauge field [49] or instantons [27] might also be considered.  These questions are currently being examined. 

\section*{Acknowledgements}
R. Macleod had useful input.

\end{document}